\documentclass[11pt]{article}
\usepackage{amsmath,amssymb}
\usepackage{authblk}
\usepackage{fullpage}
\usepackage{graphicx}
\usepackage{acronym}
\usepackage{cleveref}
\usepackage{textcomp} 
\usepackage{gensymb}
\usepackage{natbib}
\usepackage{siunitx}

\newacro{HSV}{hue, saturation, and value}
\newacro{ODE}{ordinary differential equation}
\newacro{ODIL}{optimizing with a discrete loss}
\newacro{PnP}{perspective-$n$-point}
\newacro{PLA}{polylactic acid}
\newacro{PVC}{polyvinyl chloride}
\newacro{HSV}{hue, saturation, and value}

\DeclareMathOperator*{\argmin}{arg\,min}

\newcommand{\figbf}[1]{\uppercase{\textbf{#1}}}

\graphicspath{{.},{figures/}}

\title{Contactless Precision Steering of Particles in a Fluid inside a Cube with Rotating Walls}

\author[1,$\dagger$]{Lucas Amoudruz}
\author[1,$\dagger$]{Petr Karnakov}
\author[1,$\ast$]{Petros Koumoutsakos}

\affil[1]{Computational Science and Engineering Laboratory, Harvard University, Cambridge, MA 02138, USA}
\affil[$\dagger$]{Equal contributions}
\affil[$\ast$]{corresponding author: petros@seas.harvard.edu}

\date{}

\begin{document}

\maketitle

\begin{abstract}
  Contactless manipulation of small objects is essential for biomedical and chemical applications, such as cell analysis, assisted fertilisation, and precision chemistry. Established methods, including optical, acoustic, and magnetic tweezers, are now complemented by flow control techniques that use flow-induced motion to enable precise and versatile manipulation. However, trapping multiple particles in fluid remains a challenge. This study introduces a novel control algorithm capable of steering multiple particles in flow. The system uses rotating disks to generate flow fields that transport particles to precise locations. Disk rotations are governed by a feedback control policy based on the Optimising a Discrete Loss (ODIL) framework, which combines fluid dynamics equations with path objectives into a single loss function. Our experiments, conducted in both simulations and with the physical device, demonstrate the capability of the approach to transport two beads simultaneously to predefined locations, advancing robust contactless particle manipulation for biomedical applications.
\end{abstract}

\section{Introduction}

Contactless manipulation of small objects is crucial in numerous biomedical and chemical applications, such as the manipulation and analysis of cells, bacteria, viruses, and other small particles~\citep{miyamoto2016single,guo2016three,jericho2004micro}, assisted fertilisation~\citep{medina2016cellular}, bioprinting~\citep{zhu2024voxelated}, and microreactors~\citep{teh2008droplet}.
These processes require the precise transport of one or many objects to a prescribed position, without damaging their structure.
Several techniques have been developed to achieve this task.

Optical tweezers~\citep{ashkin1986observation} focus light to create forces that trap micro- and nanoparticles.
This approach has high accuracy and has been applied successfully for particle sorting~\citep{macdonald2003microfluidic}, cell analysis~\citep{mills2004nonlinear}, and microfabrication~\citep{pauzauskie2006optical}.
However, the forces of optical tweezers are small, limiting their application to micrometre-sized and smaller objects.
Furthermore, the heat produced by the absorption of light in the medium can potentially damage biological samples or affect their mechanical temperature~\citep{liu1995evidence}.
Acoustic tweezers~\citep{wu1991acoustical} are another popular method for trapping small objects in a fluid using acoustic waves.
This approach has been used to manipulate single cells~\citep{lam2016multifunctional} and small organisms~\citep{sundvik2015effects}, and to isolate biomarkers associated with cancer detection~\citep{li2015acoustic,wu2017isolation}.
Acoustic tweezers can manipulate objects of sizes ranging from $\SI{e-7}{\meter}$ to $\SI{e-2}{\meter}$~\citep{ozcelik2018acoustic}.
However, acoustic tweezers have multiple designs that are highly specialised due to limitations related to excessive heat production in the fluid, complex setups, limited vibration patterns and limited precision~\citep{ozcelik2018acoustic}.
Other approaches for contactless manipulation include hydrodynamic forces~\citep{lutz2006hydrodynamic,van2023simple}, electrophoresis~\citep{campbell2004electrophoretic}, dielectrophoresis~\citep{wang1997dielectrophoretic}, thermocapillary flows~\citep{pinan2021microrobotic}, and magnetic forces to measure cell properties~\citep{bausch1999measurement}, manipulate droplets~\citep{li2020programmable}, and control artificial microswimmers~\citep{zhang2009artificial,amoudruz2022independent,amoudruz2024path}.
Several studies used hybrid approaches, combining the aforementioned methods to enhance precision or overcome limitations inherent in the use of each method individually~\citep{ghosh2018mobile,minzioni2017roadmap}.

In recent years, flow control techniques have gained attention for their ability to handle diverse tasks in biomedical and chemical applications.
These methods use forces generated by the flow fields to steer particles without any contact.
Fluid flow can be controlled in numerous ways, such as using pumps~\citep{xu2023compact,schneider2011algorithm}, rotational parts~\citep{ye2012micro}, and soft robotic cilia~\citep{ren2022soft}.
Examples include microfluidic devices that leverage flow patterns to sort and analyse cells~\citep{ozkumur2013inertial,karabacak2014microfluidic,miyamoto2016single,riordon2019deep}, and droplet-based systems that use micropumps to perform chemical reactions on the microscale~\citep{teh2008droplet,elvira2013past}.
A recent study~\citep{kislaya2025}, particularly relevant to this work, presented a microfluidic system for dynamic multi-particle manipulation using programmable hydrodynamics in a Hele-Shaw flow, guided by simulations.

Here, we propose a novel flow control method to transport particles towards desired configurations inside a fluid confined in a cube. The particles are transported by the velocity field induced by the vorticity produced by rotating disks at the walls of the cube.
The rotation of the disks is prescribed by a feedback control algorithm that is learnt from high-fidelity numerical simulations of the flow inside the device. Control is achieved through the \ac{ODIL} framework~\citep{karnakov2024solving,karnakov2025}, which combines a model of the system's dynamics with the path objectives into a single loss function.
The algorithms learnt in simulations are deployed in the physical device and we demonstrate its capabilities by transporting simultaneously two beads towards prescribed positions.
The present approach offers a novel solution for transporting small particles suspended in a fluid, combining the precision of feedback control with the versatility of hydrodynamic flows.
This approach potentially overcomes limitations of existing methods and provides a robust tool for various biomedical applications.

\section{Materials and Methods}

\begin{figure}
  \centering
  \includegraphics[width=0.8\textwidth]{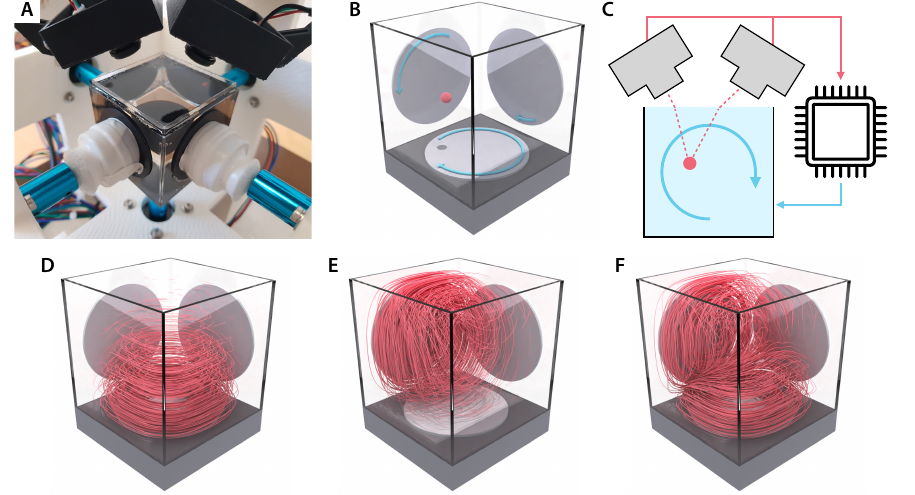}
  \caption{
    (\figbf{a})~Physical device, composed of the fluid chamber, 5 rotating disks, and 2 cameras.
    (\figbf{b})~Schematic of the device: disks rotate to create a flow that transports the beads.
    (\figbf{c})~The control policy maps the position of the beads to the disks angular velocities.
    (\figbf{d}-\figbf{f})~Flow streaklines produced by the rotation of the bottom disk (\figbf{d}), the back left disk (\figbf{e}), and both bottom and back left disks (\figbf{f}).
    The two front disks are not shown for visualisation purposes.
  }
  \label{fig:concept}
\end{figure}

The experimental setup consists of a cubic chamber, made of acrylic, of length $L=\SI{5}{\centi\meter}$, open at the top and filled with glycerol, with thin rotating disks of radius $R=\SI{2}{\centi\meter}$ mounted at the centre of each of the five inner faces.
Each disk has a thickness of $\SI{1}{\milli\meter}$ and is connected to NEMA 17 stepper motors, which are controlled by A4988 drivers linked to the PMOD interface of a Kria KR260 board.
The device is kept at room temperature.
The rotating disks generate a flow that transport suspended beads within the glycerol.
The beads have a small size of $d\approx\SI{1}{\milli\meter}$ compared to the length scale of the flow.
We test the device using beads made of either \ac{PLA}, which is neutrally buoyant, or \ac{PVC}, which sinks under gravity.
The angular velocities of the disks are controlled by a neural network-based policy, trained to guide the beads to their target positions in minimal time, based on their current locations.
Feedback for the control policy is obtained from the three-dimensional positions of the beads, which are reconstructed using two cameras positioned above the chamber and aimed at its centre.
Beads are tracked in image space by matching rectangular patches of the image that are closest to the beads colors in \ac{HSV} space.
Beads coordinates in image space are converted to rays that are refracted through the surface of the liquid, and the three-dimensional position of the beads is estimated as the closest point to the rays, in the least squares sense.
The control policy is learnt within the \ac{ODIL} framework~\citep{karnakov2024solving,karnakov2025}, which integrates a numerical model of the fluid dynamics inside the device with the path objectives into a loss function.
\Cref{fig:concept} provides an overview of the device concept and flows produced by the disks.
In the rest of this article, we express all coordinates in dimensionless units
such that the chamber size is $L=1$ and maximum angular velocity of the disks is $\omega_\text{max}=1$.


\subsection{Numerical Model}

The maximum angular velocity $\omega_\text{max}$ of the disks is chosen so that the Reynolds number of the fluid remains small.
In the experiments, we choose $\omega_\text{max} = \SI{0.45}{\radian\per\second}$, which corresponds to a Reynolds number $\mathrm{Re} = LR\omega_\text{max} / \nu \approx 0.15$, where $\nu$ is the kinematic viscosity of glycerol.
In this regime, inertial effects are negligible and we approximate the fluid's motion with the Stokes equation.
Furthermore, we assume that the top surface of the fluid remains flat and that its velocity is zero in the normal direction, but can be nonzero along its tangential axes.
We set no-slip boundary conditions on the faces of the chamber and on the disks.
By the linearity of the Stokes equation, the flow velocity when only one disk is rotating is proportional to the angular velocity of that disk.
In addition, the flow velocity created by all rotating disks is the sum of flows created by the individual disks.
The flow inside the chamber is thus given by
\begin{equation}\label{eq:flow}
  \mathbf{U}(\mathbf{x}, \mathbf{\omega}) = \sum\limits_{k=0}^{4} \frac{\omega_k}{\omega_\mathrm{ref}} \mathbf{U}_k(\mathbf{x}),
\end{equation}
where $\omega_k$ is the angular velocity of disk $k$ and $\mathbf{U}_k$ is the velocity field inside the chamber when disk $k$ has an angular velocity $\omega_\mathrm{ref}$ and disks $i \neq k$ are not rotating.
These velocity fields are numerically precomputed on a uniform grid of $128^3$ cells using a finite volume method implemented in the flow solver Aphros~\citep{karnakov2022aphros}.
Disks 0 to 4 are located on faces $x=0$, $y=0$, $x=1$, $y=1$, and $z=0$, respectively.

Assuming that beads are small compared to flow characteristics, have no inertia, and do not interact together, their positions evolve according to the \ac{ODE}
\begin{equation}\label{eq:ode}
  \dot{\mathbf{x}} = \mathbf{U}(\mathbf{x}, \mathbf{\omega}) - v_\text{sed} \mathbf{e}_z,
\end{equation}
where $\mathbf{e}_z$ is the vertical direction and $v_\text{sed}$ is the sedimentation velocity of the particle, assumed to be independent of the position of the particle.
For neutrally buoyant particles, the sedimentation velocity is zero.
In practice, the velocity fields $U_k$, $k=0,\dots,4$ in \cref{eq:flow} are stored on a grid and linearly interpolated when evaluating the right-hand side of \cref{eq:ode}.
This approach requires solving the Stokes equation only during the preprocessing stage and thus \cref{eq:ode} is not prohibitively expensive.

\subsection{Learning the control policy with ODIL}

The task consists in finding an optimal policy $\pi^\star$ that controls the disks to transport $M$ beads, with positions $\mathbf{x}_m$, $m=1,\dots, M$, from their current position to their target in minimal time:
\begin{equation} \label{eq:optimization:problem}
  \begin{aligned}
    & \pi^{\star} = \argmin\limits_{\pi} T, \\
    \text{subject to:} \quad
    & \dot{\mathbf{x}}_m = \mathbf{f}(\mathbf{x}_m, \pi(\mathbf{x}_1,\dots \mathbf{x}_M)), \quad t \in [0, T], \\
    & \mathbf{x}_m(0) = \mathbf{x}_{\text{start}, m}, \quad \mathbf{x}_m(T) = \mathbf{x}_{\text{target},m},
  \end{aligned}
\end{equation}
where $\mathbf{x}_{\text{start},m}$ and $\mathbf{x}_{\text{target},m}$ are the initial and target bead configurations, respectively, and $\mathbf{f}(\mathbf{x}, \omega) = \mathbf{U}(\mathbf{x}, \omega) - v_\text{sed} \mathbf{e}_z$ is the right hand side of \cref{eq:ode}.
The policy is represented by a neural network with weights $\theta$, $\pi \approx \pi_\theta$, and maps the beads configuration to the velocity of the disks, $\omega = \pi_\theta(\mathbf{x}_1,\dots,\mathbf{x}_M)$.
We discretize the trajectories of $\mathbf{x}$ over $N-1$ time intervals with constant time step $\Delta t = T / (N-1)$.
The constraints of \cref{eq:optimization:problem} can be enforced using a penalization on the residuals of the discretized \ac{ODE}, as described within the \ac{ODIL} framework for path planning~\citep{karnakov2025}.
Discretizing the \ac{ODE} with the midpoint rule, the penalization term for the \ac{ODE} constraint is written as
\begin{equation} \label{eq:ODIL:ODE}
  \mathcal{L}_\text{ODE}(\mathbf{x}; \theta) = \sum\limits_{m=1}^M \sum\limits_{n=0}^{N-2} \left( \Delta \mathbf{x}_m^{n+1/2} - \mathbf{f}_{\theta,m}^{n+1/2} \Delta t\right)^2,
\end{equation}
where we have defined $\Delta \mathbf{x}_m^{n+1/2} = \mathbf{x}_m^{n+1} - \mathbf{x}_m^{n}$, $\mathbf{f}_{\theta,m}^{n+1/2} = \mathbf{f}(\mathbf{x}_m^{n+1/2}, \pi_\theta(\mathbf{x}_m^{n+1/2}))$, and $\mathbf{x}_m^{n+1/2} = (\mathbf{x}_m^{n} + \mathbf{x}_m^{n+1}) / 2$.
Furthermore, the time-step is computed from the trajectory as~\citep{karnakov2025}
\begin{equation}
  \Delta t = \frac{1}{M} \sum\limits_{m=1}^{M} \frac{\Delta \mathbf{x}_m \cdot \mathbf{f}_m}{\mathbf{f}_m \cdot \mathbf{f}_m},
\end{equation}
with the scalar product $\mathbf{u}\cdot\mathbf{v} =\sum_{n=0}^{N-2}\mathbf{u}^{n+1/2}\cdot\mathbf{v}^{n+1/2}$ defined for any midpoint vectors $\mathbf{u}$ and $\mathbf{v}$.
The travel time is thus given by $T = (N-1) \Delta t$.
The \ac{ODIL} framework consists in solving both the trajectory and the objective function in a combined loss:
\begin{equation} \label{eq:odil:singletraj}
  \mathcal{L}_\text{traj}(\mathbf{x}, \theta) = \mathcal{L}_\text{ODE}(\mathbf{x}; \theta) + \lambda T,
\end{equation}
where $\lambda$ is a positive constant that controls the importance of the objective with respect to the \ac{ODE} constraints.
\Cref{eq:odil:singletraj} can be minimised with respect to the trajectory $\mathbf{x}$ and weights $\theta$ to find both an optimal policy and the corresponding beads trajectory for the given initial configuration.
However, doing so with a single instance of initial beads configuration $\mathbf{x}_{\text{start}, m}$ limits the range of validity of the policy.
Instead, we average this loss over a set of $S$ trajectories starting from different beads configurations:
\begin{equation} \label{eq:odil}
  \mathcal{L}(\mathbf{x}^{(1)}, \dots, \mathbf{x}^{(S)}, \theta) = \frac 1 S \sum\limits_{s=1}^S \mathcal{L}_\text{traj}(\mathbf{x}^{(s)}, \theta),
\end{equation}
where $\mathbf{x}^{(s)}$ is the (unknown) optimal trajectory that connects beads from initial configuration $\mathbf{x}^{(s)}_{\text{start},m}$, $s=1,\dots,S$, to the target configuration.
We thus obtain the optimal policy by minimizing \cref{eq:odil} with respect to the $S$ trajectories and the weights $\theta$.
The starting positions $\mathbf{x}_{m}^{0,(s)}$ remain constant during training, and the ending positions $\mathbf{x}_{m}^{N-1,(s)}$ of the trajectories are restricted to the target.
The initial guess for the trajectories $\mathbf{x}_{m}^{(s)}$ before optimisation is set to equidistant points on the straight line between the start and the projection of the start to the target for each bead $m=1,\dots,M$ and sample $s=1,\dots,S$.
The unknown trajectories are represented with a multi-grid decomposition (see \cref{se:multigrid}) along the time direction to accelerate convergence~\citep{karnakov2023flow}.

The optimisation was carried out with the Adam optimizer with an initial learning rate $\gamma = 0.05$ over 10\,000 epochs.
The parameter $\lambda$ and the learning rate were halved every 1\,000 epochs.
The initial value of $\lambda$ was set to $0.001$.
Furthermore, we set $N=256$ and $S=10\,000$ in all cases.
The policy is approximated by a feed-forward neural network with hidden layers of 128 units each and hyperbolic tangent activation function, with 4 and 6 hidden layers for the cases of 1 and 2 beads, respectively.
The final layer is constrained and scaled using a hyperbolic tangent function to produce values within the range $[-\omega_\text{max},\omega_\text{max}]$.
The derivatives of the loss function with respect to the discretized trajectories and weights of the policy neural network were computed through automatic differentiation with the package JAX~\citep{jax2018github}.
Training in all cases were performed on a single NVIDIA\textsuperscript{\textregistered} A100 80GB GPU and took under 10 minutes to finish.

To test the performance of trained policies, \cref{eq:ode} is numerically integrated with a second order Runge-Kutta explicit solver with the same number of time steps as used during training.
Furthermore, we use initial configurations that are randomly selected, and different from those used during training, to test the generalisation of the learnt policies.

\section{Results}

\subsection{Single bead reaching a line} \label{se:results:1bline}

\begin{figure}
  \centering
  \includegraphics[width=\textwidth]{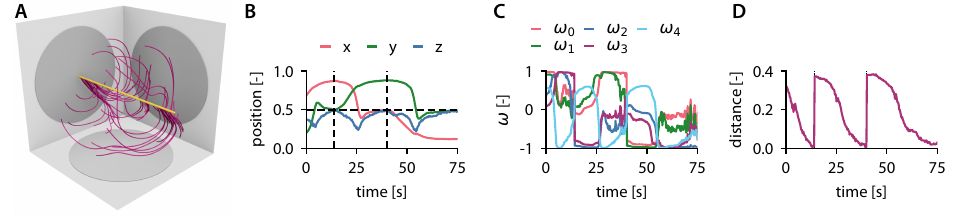}
  \caption{
    Single neutrally buoyant bead reaching a line.
    (\figbf{a})~Trajectories obtained from simulations with the trained policy, starting from 50 uniformly sampled positions and reaching the line $y=z=0.5$ (yellow).
    (\figbf{b})~Coordinates of the bead (solid lines) over time.
    The vertical dashed lines correspond to a switch of target between the lines $x=z=0.5$ and $y=z=0.5$.
    (\figbf{c})~Angular velocity of the five rotating disks.
    (\figbf{d})~Distance between the bead and its target against time.
    Vertical dashed lines correspond to times where the target is rotated by $90\degree$ with respect to the vertical axis.
  }
  \label{fig:1b:lines}
\end{figure}

We first consider a neutrally buoyant bead starting from a random position, uniformly sampled within the chamber, with the task of reaching a specific line.
We select the line $y = z = 0.5$, meaning that only the $x$ component of $\mathbf{x}^{N-1,(s)}$ is subject to optimisation.
The sedimentation velocity $v_\text{sed}$ is set to $0$ in the numerical model.
The trained policy is tested using the numerical model for multiple random starting positions.
\Cref{fig:1b:lines}\figbf{a} shows simulated trajectories following the trained policy.
These trajectories start from 50 random positions uniformly sampled within the chamber and successfully converge to the target line.
We then apply the control policy learnt from numerical simulations to the physical device by placing a neutrally buoyant bead at a random position inside the chamber.
The feedback control adjusts the angular velocities of the disks to guide the bead to the target line.
Once the bead reaches its target, the target is switched to the line $x=z=0.5$, and after the bead reaches this new line, the target reverts to the original line.
\Cref{fig:1b:lines} shows the trajectory of the bead, the angular velocities of the disks, and the distance of the bead to the current target, over time.
The device consistently brings the bead within a small distance to its target in about $\SI{25}{\second}$.
This time is limited by the angular speed $\omega_\mathrm{max}$, that must be sufficiently low to remain in the Stokes regime.
The angular velocities follow a complex and noisy pattern over time, reflecting the complexity of the trajectories and the noisy feedback from the vision system.
We note that in addition to the noise in the spatial measurements, there is a time delay in the control loop due to the capture and processing of the camera images.
However, the temporal resolution is typically sufficient as the displacement of the bead between two frames remains in the order of $\SI{0.5}{\milli\meter}$, a distance smaller than the bead size.

\begin{figure}
  \centering
  \includegraphics[width=0.3\textwidth]{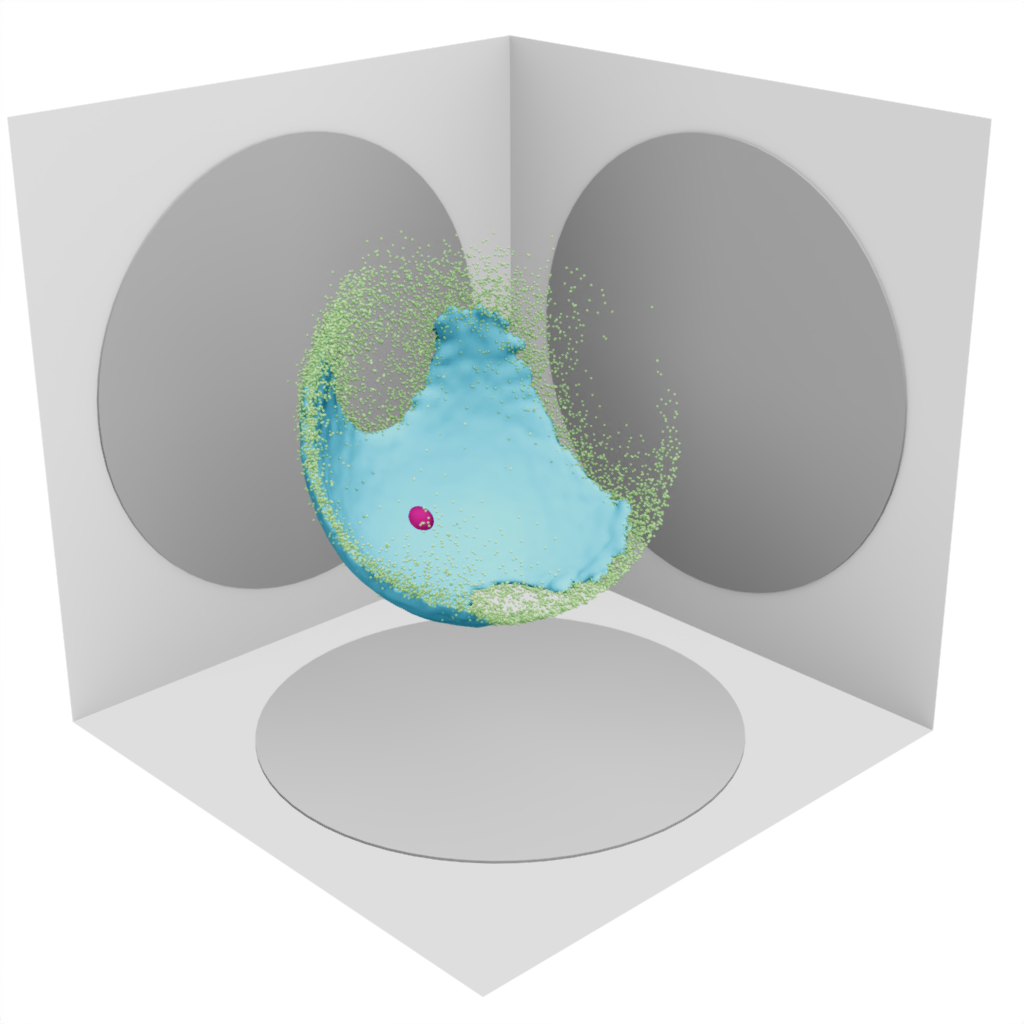}
  \caption{Final positions of a neutrally buoyant bead following a random policy.
  The bead starts at $(0.3, 0.5, 0.3)$ (purple dot).
  All five disks rotate with random angular velocities sampled from
  $\mathcal{U}[-1,1]$ over 200 steps, each with duration of $0.2\pi$, corresponding to an upper bound of 20 full rotations.
  Final positions from 100\,000 samples (green dots) and an isosurface of their number density (blue surface) remain close to a two-dimensional manifold.
  }
  \label{fig:limitation:buoyant}
\end{figure}

We remark that, for this specific task, we have selected a line instead of a point target due to a limitation of the device for neutrally buoyant beads.
Indeed, neutrally buoyant beads remain on a two-dimensional manifold, as shown in \cref{fig:limitation:buoyant}.
This is due to the symmetry of the device: Each disk can translate one bead on a circular loop centred on the axis of rotation of the disk.
Thus, the reachable positions for a passive tracer remain on a spheroid centred around the middle of the chamber.
This limitation prevents neutrally buoyant beads from moving along the radial direction.
Nevertheless, lines passing through the centre of the chamber are therefore reachable from any initial position of the bead, hence the choice of the target line $y = z = 0.5$.
In the following sections, we will demonstrate that dense beads can escape these manifolds thanks to gravity.

\subsection{Single sinking bead reaching a point}

\begin{figure}
  \centering
  \includegraphics[width=\textwidth]{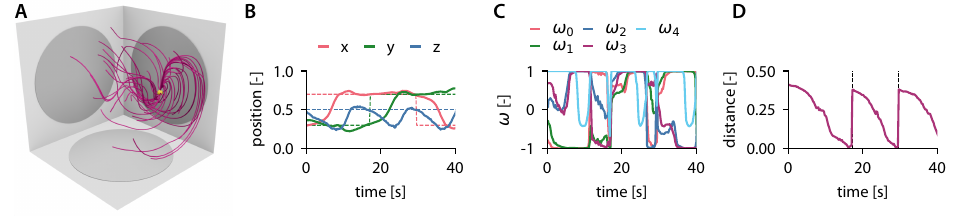}
  \caption{
    Single bead with a higher density than the fluid, reaching a point.
    (\figbf{a})~Trajectories obtained from simulations with the trained policy, starting from 50 uniformly sampled positions and ending at the target $(0.7, 0.7, 0.5)$ (yellow).
    (\figbf{b})~Coordinates of the bead (solid lines) and their target (dashed lines) over time.
    (\figbf{c})~Angular velocity of the five rotating disks.
    (\figbf{d})~Distance between the bead and its target against time.
    Vertical dashed lines correspond to times where the target is rotated by $90\degree$ with respect to the vertical axis.
  }
  \label{fig:1b:gravity:point}
\end{figure}

Next, we consider a single bead with a density higher than that of glycerol.
Under these conditions, the bead sinks at a constant speed due to gravity.
We ignore wall effects and assume that the sedimentation velocity is constant throughout the chamber.
This sinking motion provides an additional degree of freedom for transporting the bead to its target.
A new control policy is trained to guide the bead to the point $(0.7, 0.7, 0.5)$.
The trained policy is tested on random starting positions.
The policy reliably guides the bead to its target, as demonstrated on 50 trajectories starting from uniformly sampled positions within the chamber (\cref{fig:1b:gravity:point}\figbf{a}).

The policy that was trained on numerical simulations is then tested on the physical device, starting with a single bead placed randomly inside the chamber.
The motors follow the control policy, guiding the bead to its target, as demonstrated in \cref{fig:1b:gravity:point}.
After the bead reaches the target, the target position is rotated by $90\degree$ around the vertical axis $x=y=0.5$.
The device consistently guides the bead to its target within a negligible distance compared to the size of the chamber.
The policy generates a complex sequence of angular velocities for the disks (\cref{fig:1b:gravity:point}), with angular velocities often saturating to their maximum allowed value.
This suggests that the policy minimises the travel time to the target.

\subsection{Manipulating two sinking beads}

\begin{figure}
  \centering
  \includegraphics[width=0.7\textwidth]{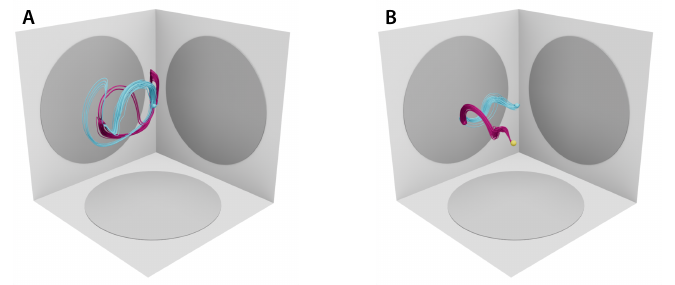}
  \caption{
    Trajectories of two beads following the trained policy in simulations.
    (\figbf{a})~Swapping positions.
    (\figbf{b})~Rotating positions around the line $x=y=0.5$ by $90\degree$.
    Blue and purple spheres denote the positions $\mathbf{x}_{1,s}$ and $\mathbf{x}_{2,s}$, respectively, and the yellow sphere in (\figbf{b}) shows the new target of the second bead.
  }
  \label{fig:two:traj}
\end{figure}

We now test the device on three tasks that involve the manipulation of two beads that have a higher density than the fluid: trapping beads, swapping beads, and rotating bead configurations.
A single control policy is used for all tasks, trained to bring each bead to its respective target position.
During training, the beads start from random positions, uniformly distributed within the chamber, with targets set to $\mathbf{x}_{1}^{N-1,(s)} = (0.3, 0.3, 0.5)$ and $\mathbf{x}_{2}^{N-1,(s)} = (0.3, 0.7, 0.5)$.
The policy is first tested on the numerical model, and we consider two tasks: swapping beads and rotating their configuration.
The initial positions of the beads are randomly sampled from a normal distribution centred around $\mathbf{x}_{1}^{N-1,(s)}$ and $\mathbf{x}_{2}^{N-1,(s)}$ with a small standard deviation $\sigma = 0.008$.
The targets are either swapped or rotated according to the task, and the simulated trajectories of the beads are shown on \cref{fig:two:traj}.
In both cases, the policy successfully guides the beads to their new target.

\begin{figure}
  \centering
  \includegraphics[width=\textwidth]{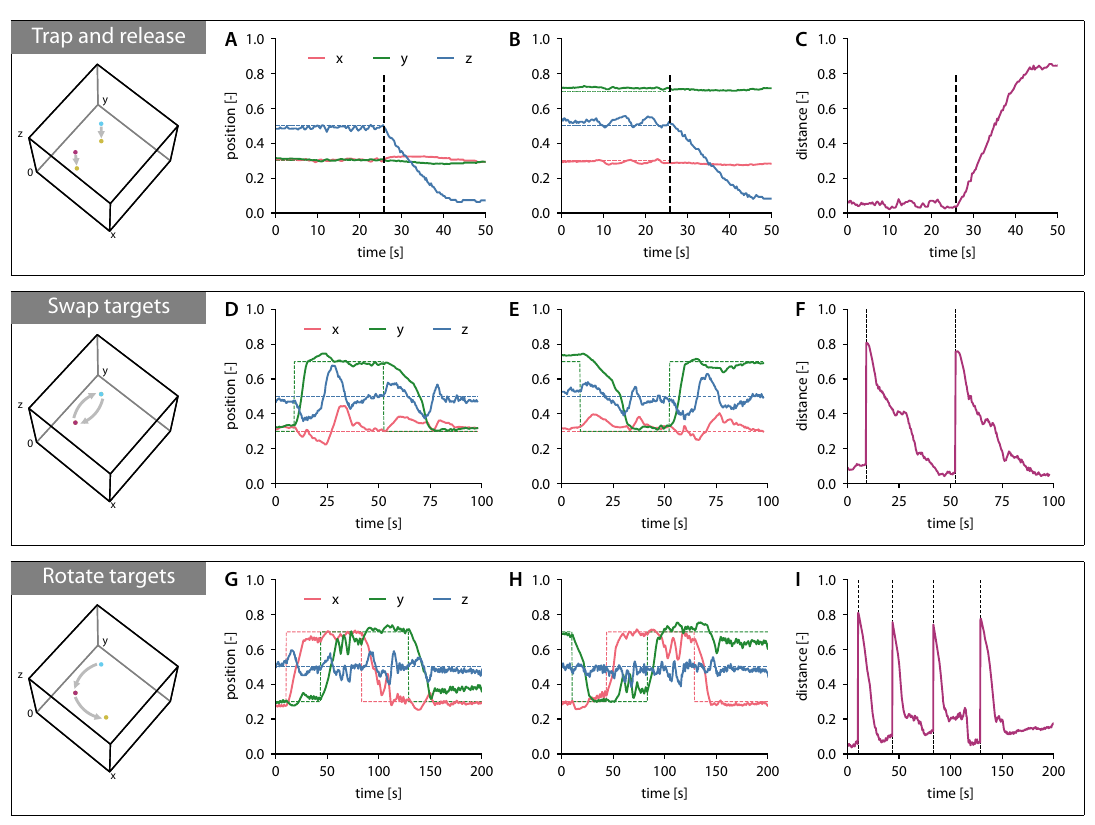}
  \caption{
    Physical device performing three tasks: trapping beads at prescribed locations (\figbf{a}-\figbf{c}), swapping beads (\figbf{d}-\figbf{f}), and rotating the bead configuration (\figbf{g}-\figbf{i}).
    (\figbf{a},\figbf{d},\figbf{g})~Coordinates of the first bead against time.
    (\figbf{b},\figbf{e},\figbf{h})~Coordinates of the second bead against time.
    (\figbf{c},\figbf{f},\figbf{i})~Distance to target against time.
    The vertical dashed lines indicate key events for each task: end of trapping, target swapping, and targets rotation, respectively.
  }
  \label{fig:two:tasks}
\end{figure}

This same policy is then tested with the physical device in the trapping task.
When the policy is active, the beads move towards their target position, and the device successfully traps the beads at the prescribed locations (\cref{fig:two:tasks} and Movie 1).
When the motors are at rest, or after releasing the trap, the beads sink to the bottom of the chamber ($z=0$).
As predicted by Stokes' law, and neglecting wall effects, the beads sink at a constant velocity.

Next, we test the device for the task of swapping beads.
This is achieved by swapping the two targets after the beads reach their respective targets.
To reuse the same policy, we simply swap the positions in the input of the control policy.
The resulting bead trajectories are shown in \cref{fig:two:tasks} and in Movie 2.
After each target swap, the beads successfully reach their new targets within a distance that is below 10\% of the chamber dimensions, with a transition time of about $\SI{30}{\second}$.

Finally, we test the device on the rotation task.
This task consists in rotating the targets around the line $x=y=0.5$ by increments of $90\degree$, every time the beads reach their targets.
By rotating the coordinate system and adjusting the policy output accordingly, the same policy is used for this task as well.
The trajectories of the beads and the distance to the target are shown in \cref{fig:two:tasks} and in Movie 3.
Once again, the device successfully completes the task, guiding the beads to their desired positions within a distance that is $10\%$ of the box length in the range of $\SI{50}{\second}$.

\subsection{Manipulation of three or more beads}

\begin{figure}
  \centering
  \includegraphics[width=0.3\textwidth]{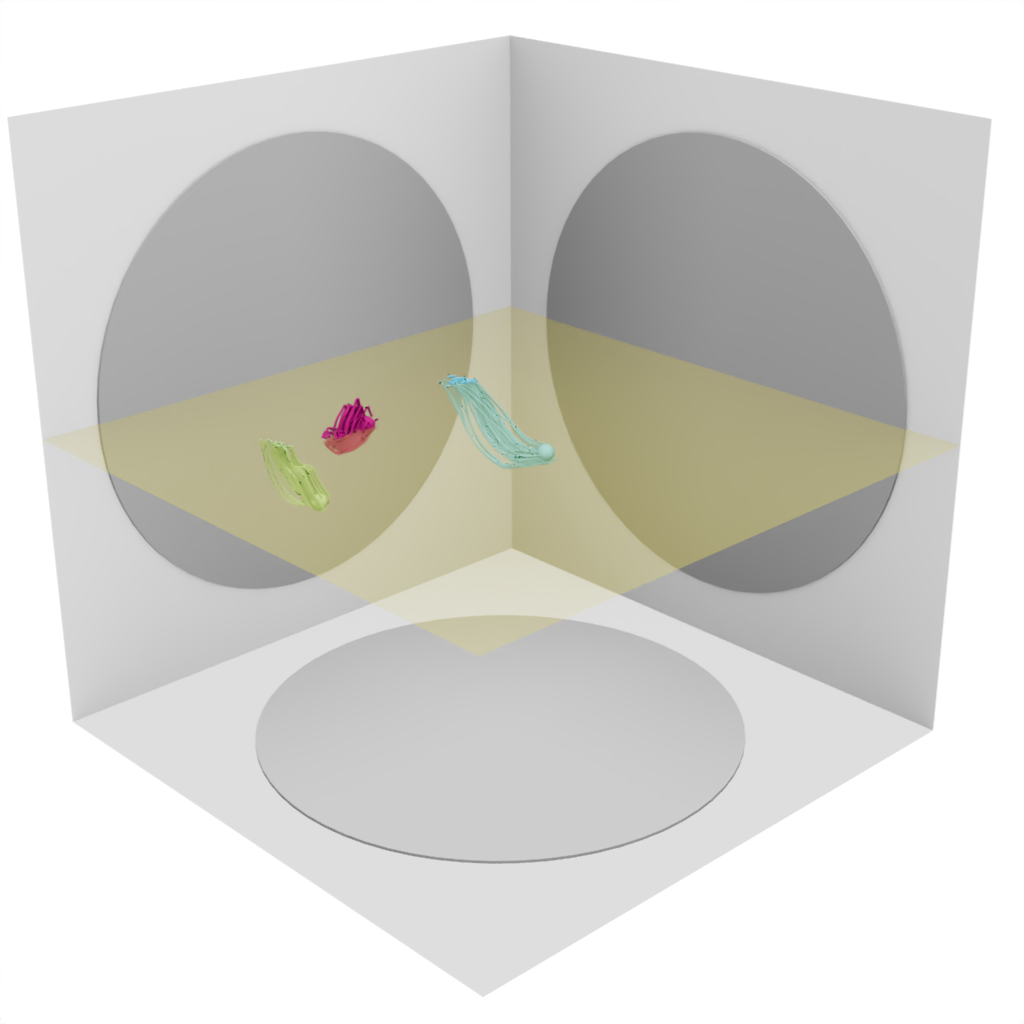}
  \caption{Three buoyant beads reaching a plane (yellow) when following a policy trained for this task.}
  \label{fig:3b:plane}
\end{figure}

So far we have studied the control of a maximum of two beads, for which the five control inputs, in addition to gravity, were sufficient to control the 6 degrees of freedom of 2 beads.
Similarly, it is possible to guide three beads to prescribed planes, as shown in \cref{fig:3b:plane}.
The policy was trained with 40\,000 trajectories, with three beads initially uniformly distributed in $[0.25, 0.75]^3$, with the plane target $z=0.5$.
\Cref{fig:3b:plane} show the trajectories of 3 beads starting at normally distributed positions around $(0.3, 0.4, 0.5)$, $(0.4, 0.7, 0.4)$ and $(0.3, 0.3, 0.4)$, with a standard deviation $\sigma = 0.008$.
All beads reach the target plane.
We note that with this specific device, we were not able to learn successful policies that precisely control the path of more than three beads.
This is attributed to the five disks that are available as controllers.
Two-dimensional simulations in \citet{karnakov2025} have shown that controlling more beads would be possible with more rotating disks.
We also note that the recent work \citep{kislaya2025} argued that $2N+1$ controllers are needed for precise manipulation of $N$ particles in two dimensions.
The controllability of small particles has been studied under similar conditions by \citet{or2009geometric,walker2022control} and will be the object of further studies in the case of rotating disks.

\section{Summary}

We have introduced a novel approach for transporting small objects suspended in a fluid, which combines the precision of feedback control with the versatility of hydrodynamic flows. We have demonstrated the capabilities of the approach in simulations and in a corresponding experimental device.
The device is controlled by a policy learnt through the \ac{ODIL} framework, and we have demonstrated its ability to transport one or two beads simultaneously to a target.
The device can manipulate both neutrally buoyant beads and beads with a density higher than that of the fluid, which would otherwise sink in the absence of control.

The control algorithm plays a crucial role in the successful operation of the device.
We have found that the policy is robust to measurement and modelling errors, as it was trained on 10\,000 simulated trajectories.
We anticipate that this approach can be transferred to other physical systems, opening possibilities for controlling devices that rely on multiple physical forces simultaneously.
This method is a potent approach for manipulating particles in biomedical and chemical applications at the microscale, where contactless manipulation is often essential.

\subsection*{Supplementary data}
Supplementary movies are available at https://doi.org/10.1017/jfm.2025.10174.

\subsection*{Acknowledgements}
The authors acknowledge the support of MassRobotics, Dassault Systemes, igus, and AMD in providing hardware parts and software for building the prototype.

\subsection*{Declaration of interests}
The authors report no conflict of interest.

\section*{Appendix}
\appendix

\section{Multigrid decomposition}
\label{se:multigrid}

The unknown discretised trajectories are represented with a multigrid decomposition technique to accelerate the convergence of ODIL~\citep{karnakov2023flow}.
We consider a hierarchy of successively coarser grids of size $N_i = N / 2^{i-1}$ cells for $i=1,\dots,L$, where $L$ is the total number of levels.
The multigrid decomposition operator is defined as
\begin{equation}
M_L(u_1, \dots, u_L) = u_1 + T_1 u_2 + \dots +  T_1 T_2 \dots T_{L-1} u_{L},
\end{equation}
where each $u_i$ is a field on a grid with size $N_i$,
and each $T_i$ is the interpolation operator from the grid of size $N_{i+1}$ to the finer grid of size $N_i$.
The multigrid decomposition of a discrete field $u$ on a grid of size $N$ is thus
\begin{equation}
u = M_L(u_1, \dots, u_L).
\end{equation}
We remark that this representation is over-parametrised and therefore not unique.
The total number of scalar parameters increases from $N^d$ of the original field $u$ to $N_1^d + \dots + N_L^d$ for the representation $u_1,\dots,u_L$, where $d$ is the dimension of the grid.
In this study, $d=1$ since we apply the multigrid decomposition only along time.
The interpolation operators~$T_i$ are defined from linear interpolation~\citep{trottenberg2000multigrid} for node-based discretisation.
As an illustration, consider a hierarchy of grids in one dimension with $N = N_1 = 8$, $N_2 = 4$, and $N_3 = 2$ cells.
The node-based interpolation matrices $T_1 \in \mathbb{R}^{9\times 5}$ and $T_2 \in \mathbb{R}^{5\times 3}$ are
\[
T_1 = \frac{1}{2}
\begin{bmatrix}
2 & 0 & 0 & 0 & 0 \\
1 & 1 & 0 & 0 & 0 \\
0 & 2 & 0 & 0 & 0 \\
0 & 1 & 1 & 0 & 0 \\
0 & 0 & 2 & 0 & 0 \\
0 & 0 & 1 & 1 & 0 \\
0 & 0 & 0 & 2 & 0 \\
0 & 0 & 0 & 1 & 1 \\
0 & 0 & 0 & 0 & 2 \\
\end{bmatrix}, \quad
T_2 = \frac{1}{2}
\begin{bmatrix}
2 & 0 & 0 \\
1 & 1 & 0 \\
0 & 2 & 0 \\
0 & 1 & 1 \\
0 & 0 & 2 \\
\end{bmatrix}.
\]
In this work, we used the node-centred interpolation matrices along time.
This technique addresses the issue of locality of gradient-based optimisers by extending the domain of dependence of each scalar parameter so that information can propagate through the grid faster.

\bibliography{refs}
\bibliographystyle{unsrt}

\end{document}